\documentclass[fleqn,usenatbib]{mnras}
\usepackage{newtxtext,newtxmath}

\usepackage{graphicx}	
\usepackage{amsmath}	
\usepackage{amssymb}	
\usepackage{tabularx}
\usepackage[flushleft]{threeparttable}
\usepackage[a4paper]{geometry}%
\usepackage{hyperref}

\title[Geminga pulsar and pulsar wind nebula]{Positron flux and gamma-ray emission from Geminga pulsar and pulsar wind nebula}

\author[Xiaping Tang, Tsvi Piran]{
Xiaping Tang,$^{1}$\thanks{E-mail: tang.xiaping@mail.huji.ac.il, tangxiaping@gmail.com}
Tsvi Piran,$^1$
\\
$^{1}$ The Racah Institute of physics, The Hebrew University of Jerusalem, Jerusalem 91904, Israel\\\
}

\date{Accepted XXX. Received YYY; in original form ZZZ}

\pubyear{2018}

\begin{document}
\label{firstpage}
\pagerange{\pageref{firstpage}--\pageref{lastpage}}
\maketitle

\begin{abstract}
Nearby pulsars have been suggested as sources of $\sim$TeV $e^+/e^-$ Cosmic Ray (CR) excess on Earth. The High-Altitude Water Cherenkov Observatory (HAWC) detected extended TeV emission regions in the direction of two nearby middle-aged pulsars, Geminga and PSR B0656+14. By modeling the TeV emission as inverse Compton emission from electron-positron pairs diffusing in the interstellar medium (ISM), the HAWC collaboration derives a diffusion coefficient much smaller than the standard value in the vicinity of the two pulsars, which make them unlikely the origin of the positron excess. We propose that the observed $\gamma$-ray emission originate from the relic pulsar wind nebula. A two zone diffusion model with a slow diffusion in the nebula and a fast diffusion in the ISM can explain the HAWC surface brightness profile and the positron excess simultaneously. Inefficient diffusion in the $\gamma$-ray emission region surrounding a middle-aged pulsar maybe a common phenomenon that can be tested by future observation. The implied diffusion coefficient in the ISM is smaller than the one suggested by the standard CR propagation model, but it is fully consistent with the predictions of the spiral arm model.

\end{abstract}


\section{Introduction}
The origin of cosmic ray (CR) is a long standing problem. CRs are composed of primarily protons and atomic nuclei with a small fraction of positrons and electrons. CR particles with energy below the so-called CR knee ($10^{15}$eV) are confined within our Galaxy by the Galactic magnetic fileds, thus they must have a Galactic origin. CR particles with energy above the CR knee are instead considered to be extra-galactic. The standard picture for the origin of Galactic CR assumes that the CR particles are accelerated at Galactic supernova remnant shocks and then diffuse to Earth. 

Recently, {\it PAMELA} \citep{Adriani10} and {\it AMS-02} \citep{Aguilar13} detected an excess of positron flux at energies from tens of GeV to hundreds of GeV compared with the prediction of the standard picture \citep{Strong07}. Both dark matter particle interaction \citep[eg.,][]{Ibarra14} and inhomogeneity of astronomical sources \citep[eg.,][]{Hooper09,Shaviv09} have been proposed to explain the observed positron excess. \cite{Hooper09} attribute the positron excess purely to the pulsars in our Galaxy as they are good sources of positron. \cite{Shaviv09} show that the positron excess below $\sim 300$GeV can also be explained by a spiral arm model, in which the supernova rate is higher in the spiral arm region than in the inter-arm region due to the higher concentration of supernova remnants.  
In the spiral arm model, the pulsar contribution is important to the positron excess at $\gtrsim 300$GeV, which requires nearby pulsars located within a few hundreds of pc from the Earth. It worth noting that the spiral arm model can not only explain the positron excess but also other CR anomaly like the rising spectrum of CR below 1GeV, the boron to carbon ratio \citep{Ben14} and sub-Fe/Fe ratio \citep{Ben16}. Therefore, unless specifically noted, in the following discussion we focus on the positron excess $\gtrsim 300$GeV. Two promising candidate pulsars for the positron excess above $\sim 300$GeV are Geminga and PSR B0656+14 with distances of 250pc and 288pc respectively \citep{Brisken03,Verbiest12}. 

Energetic electron-positron pairs that escape from the pulsar produce inverse Compton (IC) and synchrotron $\gamma$-ray through interaction with radiation and magnetic field in the vicinity of the pulsar. Extended emission regions in the direction of Geminga and PSR B0656+14 pulsar \citep{Abe17} have been revealed recently by {\it HAWC} between 1 and 50TeV.  By modeling the TeV emission as IC emission from electron-positron pairs diffusing in the ISM, the {\it HAWC} collaboration derives a diffusion coefficient of $D\approx (4.5\pm 1.2) \times 10^{27}\rm cm^2/s$ at $100$TeV using a joint fit of the surface brightness profiles of Geminga and PSR B0656+14. This fitted $D$ value in the $\gamma$-ray emisison region is much smaller than the standard value of CR diffusion coefficient in the ISM, which is $D_{ISM}\sim 10^{28}\rm cm^2/s$ at 1GeV with a rigidity dependence of 0.3-0.6 \citep{Strong07}. Assuming such a small $D$ in the ISM between the pulsar and the Earth, \cite{Abe17} argue that the positron flux from Geminga pulsar contributes only a few percents of the observed positron excess, while the positron flux from PSR B0656+14 is negligible. 


In this work, we propose that the $\gamma$-ray emission detected by {\it HAWC} is originated from the relic pulsar wind nebula (PWN). Then the $HAWC$ surface brightness profile and the positron excess can be explained simultaneously by a two zone diffusion model with a slow diffusion in the relic PWN and a fast diffusion in the ISM. In this paper, we focus on Geminga pulsar and its PWN, but note that PSR B0656+14 is also potentially important for the positron excess. As according to the X-ray morphology PSR B0656+14 is possibly receding from us at a velocity $\gtrsim 400 \rm km/s$ \citep{Birzan16}. Since the pairs flux reaching Earth around 1TeV have been injected at an early phase when the pulsar was closer to us, the positron flux from PSR B0656+14 is expected to be enhanced after we take the recession into account. 

In section \ref{sec:observation}, we describe the observations. Our model for the Geminga pulsar and nebula is presented in section \ref{sec:model}. In section \ref{sec:IC_emission}, we compare the calculated IC emission of the Geminga nebula with the $\gamma$-ray data. In section \ref{sec:spatial_distribution}, we compare the two zone diffusion model results with the $HAWC$ surface brightness profile and the positron excess data.
We discuss the implication of our results in section \ref{sec:discussion}.

\section{Observational properties}{\label{sec:observation}}
\subsection{Positron excess}
In the standard picture of CR production and propagation, positrons are produced when the CR nuclei interact with interstellar medium. Within the standard CR model\footnote{Note however that in the standard model ad-hoc re-acceleration or winds are needed to explain the very low energy behavior \citep{Strong07}.}, that assumes an axisymmetric galactic density profile, the positron fraction is expected to decrease with energy, which is found to be correct below $10$GeV. Above $10$GeV, $PAMELA$ measured a positron fraction that increases with energy \citep{Adriani10}. This was later confirmed by {\it AMS-02} \citep{Aguilar13} and $Fermi$ \citep{Ackermann12}. From 10 to 500GeV, the positron flux $E^3\phi_{e^+}(E)$ appears to gradually increase from 10 to 20 $\rm GeV^2m^{-2}sr^{-1}s^{-1}$, while the spectral index of $\phi_{e^+}(E)$ varies from $−2.97\pm 0.03$ to $2.75\pm 0.05$ between 10 and 200GeV \citep{Aguilar14}.

\subsection{The Geminga pulsar}\label{sec:observation_pulsar}
The radio-quiet $\gamma$-ray pulsar Geminga (PSR J0633+1746) has a period of
$P = 0.237s$, first discovered by {\it ROSAT} in X-ray observation \citep{HH92}, and a period derivative of $\dot{P}=11.4\times 10^{-15}s\,s^{-1}$, first measured by {\it Compton Gamma Ray Observatory} in $\gamma$-rays \citep{Bertsch92}. Theses period and period derivative indicates a characteristic age of \citep{GP06}
\begin{equation}
\tau_{c}=\frac{P}{2\dot{P}}\approx 330 \rm kyr
\end{equation} 
and a spin down power of 
\begin{eqnarray}
\dot{E}_{spin}&=&4\pi^2 I \frac{\dot{P}}{P^3} \approx 3.4 \times 10^{34} \rm erg/s \nonumber\\
&\times &\left(\frac{I}{10^{45}g\, cm^2}\right)\left(\frac{\dot{P}}{11.4\times 10^{-15}s\,s^{-1}}\right)\left(\frac{0.237}{P}\right)^3,
\end{eqnarray} 
where $I$ is the neutron star's moment of inertia. 

The distance of the Geminga pulsar is 
$d=250\substack{+230\\-80}$pc after correction of the Lutz--Kelker bias \citep{Verbiest12}.
\cite{Faherty07} measured a proper motion of
$178.2 \pm 1.8$ mas/yr, which corresponds to a
transverse velocity of $ v_t \approx 211 (d/250\rm pc) \rm \, km s^{-1}$. Since the transverse velocity is much larger than the typical ISM sound speed, a bow-shock-tail PWN around Geminga is expected. 

\subsection{The Geminga PWN}
A bow-shock-tail nebula surrounding the Geminga pulsar is detected in X-ray by XMM-Newton \citep{Caraveo03} and Chandra \citep{Posselt17}. It is produced by the synchrotron emission of the electron-positron pairs in the nebula. The X-ray nebula is characterized by two long lateral tails $\approx 2'$ and an segmented axial tail $\approx 45$", whose origin is still under debate \citep{Posselt17}. 

In $\gamma$-ray, {\it MILAGRO} and {\it HAWC} revealeded an extended emission in the TeV band around the Geminga pulsar. {\it MAGIC} observed the region around the pulsar and provides an upper limit at $50$GeV. Since the TeV emission region is larger than $MAGIC$'s field of view, it is not straightforward to compare the $MAGIC$ upper limit with the flux measured by {\it MILAGRO} and {\it HAWC}. Assuming the $MAGIC$ upper limit corresponds to the emission from a circular region with an angular diameter of $3.5^\circ$, we can estimate the corrected upper limit for the entire emission region by using the approximate surface brightness formula given in eq. (1) of \cite{Abe17}. The $\gamma$-ray data are summarized in Table \ref{table:PWN_flux}. 

Because the $\gamma$-ray emission region is much larger than the X-ray nebula, \cite{Abe17} attributed the $\gamma$-ray emission to IC emission of the electron-positron pairs diffusing in the ISM. Here, we propose that both the X-ray and $\gamma$-ray emission originate from the PWN of the Geminga pulsar (see section \ref{sec:model}).

\begin{table*}
\centering
\caption{$\gamma$-ray flux of Geminga PWN}
\begin{threeparttable}
\begin{tabular}{ccccc}
\hline\hline
Instrument & photon energy (TeV) & $\gamma$-ray flux ($\rm TeVs^{-1}cm^{-2}$) & angular FWHM (degree) & reference\\
\hline
MILAGRO&$35$ &$(4.62\pm 1.31) \times 10^{-13}$  &$2.6^{+0.7}_{-0.9}$ & \cite{Abdo09}\\
HAWC & 7 &$(2.39\pm 0.34)\times 10^{-12}$ & $\sim 5$ & \cite{Abe17b} \\
MAGIC & 0.05 &$ <3.5 \times 10^{-12} $ & -- & \cite{Ahnen16} \\
MAGIC$^*$ (corrected)& 0.05 &$ < 2.6 \times 10^{-11}$ & -- & -- \\
\hline\hline
\end{tabular} 
$^*$The corrected upper limit is derived by adopting the surface brightness profile provided in eq. (1) of \cite{Abe17}.
\end{threeparttable}
\label{table:PWN_flux}
\end{table*}

\section{The Model of Geminga pulsar and its PWN}\label{sec:model}
Our model for Geminga involves a dynamically evolved pulsar and a static nebula. The main caveat of the current model is that for simplicity we neglect the spatial and time evolution of the PWN. We discuss this issue in section \ref{sec:discussion}.

\subsection{Pulsar}
We assume that the pulsar spin down energy is dissipated via magnetic dipole radiation with a braking index of $n=3$. The current age of the Geminga pulsar is estimated to be
\begin{equation}
t_{age}=\tau_c\left[1-\left(\frac{P_0}{P}\right)^2\right]\approx 320 \rm kyr,
\end{equation}
where we assumed an initial period of $P_0=0.045s$ following \cite{Abe17}.
$\tau_0$ is the initial spin down time scale:
\begin{equation}
\tau_0 = \tau_{c}\left(\frac{P_0}{P}\right)^2\approx 12\,{\rm kyr}\left(\frac{P_0}{0.045s}\right)^2.
\label{eq:initial_spindown_time}
\end{equation}
The spin down luminosity $L(t)$ is modeled as \citep[e.g.,][]{ST83}
\begin{equation}
L(t)=\dot{E}_{spin}\frac{(1+t_{age}/\tau_0)^2}{(1+t/\tau_0)^2},
\label{eq:luminosity_evolution}
\end{equation}
where $t$ is the time since the birth of the Geminga pulsar.
The basic parameters of the Geminga pulsar are summarized in Table \ref{table:Geminga}.

We denote by $\eta$ the fraction of the pulsar spin down luminosity that accelerates the electron-positron pairs in the PWN: 
\begin{equation}
\eta L(t) = 2\int_{E_{lo}}^{E_{hi}} E\,Q(E,t)dE.
\end{equation}
$Q(E,t)$ represents the differential number density of the injected electron or positron at energy $E$ and time $t$. $E_{lo}$ and $E_{hi}$ are the low energy and high energy cutoffs of the injected spectrum respectively. Following \cite{Abe17}, we adopt $E_{lo}=1\rm GeV$ and $E_{hi}=500 \rm TeV$.

The factor $\eta$ cannot be derived from first principles, as particle acceleration in PWN is not fully understood yet. It is also not clear whether $\eta$ evolves with time. For simplicity, we assume that $\eta$ is time independent. We discuss later in section \ref{sec:discussion} how would a time dependent $\eta$ affect our conclusion. 

In order to explain the positron excess, a minimum requirement without taking into account the energy loss and propagation effect is that the energy of the accelerated positrons from all the nearby pulsars together is larger than the observed positron excess, i.e., 
\begin{equation}
\sum_i \eta L_i(t=0)\left( \frac{d_i}{250\rm pc}\right)^{-2}\gtrsim 2\times 10^{36} \rm erg/s. 
\end{equation}
$t=0$ indicates that it is the initial spin down luminosity. This constrains the initial period of the nearby pulsars. For example, if we attribute the positron excess solely to the Geminga pulsar, then the simple argument requires $P_0\lesssim  0.1\rm s \,\eta^{1/4}$. 

The injected spectrum is assumed to have a broken power law with
\begin{equation}
Q(E, t_0) = q(t_0)\begin{cases}
\left(\frac{E}{E_{b}}\right)^{-\alpha_1}, \mbox{ for } E_{lo}<E<E_{b},\\
\left(\frac{E}{E_{b}}\right)^{-\alpha_2}, \mbox{ for } E_b \leq E <E_{hi},\\
\end{cases}
\label{eq:source_function}
\end{equation}
where $E_b$ is the break energy, $\alpha_1$ and $\alpha_2$ are the power law indices below and above the break respectively and $q(t_0)$ is a normalization constant:
\begin{equation}
q(t_0)=\frac{\eta L(t_0)}{2}\left[ \frac{E_b^{\alpha_1}(E_b^{2-\alpha_1}-E_{lo}^{2-\alpha_1})}{2-\alpha_1}+ \frac{E_b^{\alpha_2}(E_{hi}^{2-\alpha_2}-E_{b}^{2-\alpha_2})}{2-\alpha_2}\right]^{-1}
\end{equation} 
for $\alpha_1\neq 2$ and $\alpha_2\neq 2$. We adopt $\alpha_1=1.5$ for simplicity and $\alpha_2=2.34$ as suggested by \cite{Abe17}. If $\alpha_1$ is smaller, then the injection spectrum at low energy is too hard. If $\alpha_1$ is larger and close to 2, then it is difficult to explain the {\it MAGIC} upper limit as discussed later.

\begin{table}
\centering
\caption{Physical parameter of the Geminga pulsar}
\begin{threeparttable}
\begin{tabular}{lll}
\hline\hline
Period& $P$& 0.237s  \\
Derivative of period &$\dot{P}$&$11.4\times 10^{-15}s\,s^{-1}$\\
Initial period & $P_0$ & 0.045s\\
Spin down power & $\dot{E}_{spin}$ & $3.4 \times 10^{34}\,$ erg/s\\
Characteristic age &$\tau_c$ & $330\,$kyr \\
Age &$t_{age}$ & $320\,$kyr \\
Distance &$d$ & $250\,$pc \\
Transverse velocity &$v_t$ & $211\, \rm km s^{-1}$\\
\hline\hline
\end{tabular} 
\end{threeparttable}
\label{table:Geminga}
\end{table}
\subsection{PWN}
The rapidly rotating pulsar drives a relativistic and magnetized wind into the surrounding medium. The interaction inflates a bubble of energetic particles and magnetic fields, which is referred to as the PWN and is usually observable in multi-wavelength \citep{GP06}. The best known example of PWN is the young Crab nebula that has a prominent jet-torus structure \citep{Hester08}. In the late time evolution of a PWN, the reverse shock of the supernova remnant eventually moves inward and crushes the central PWN. This process further complicates the multi-wavelength morphology. Due to the pulsar motion and the asymmetry in the supernova ejecta and the ISM, after the crushing phase the PWN is likely to end up with two distinct parts: a compact nebula near the pulsar filled with recently injected fresh particles and an offset relic nebula dominated by aged particles injected a long time ago \citep{GP06,KRP13}. The relic PWNe can produce TeV emission via IC scattering of the interstellar radiation field and are the dominant population of $\gamma$-ray sources in our Galaxy \citep{KRP13}. Many of the compact PWNe show bow-shock-tail morphology as the central pulsar are usually found to move supersonically in the ISM.

\begin{figure}
\begin{center}
\includegraphics[width=\columnwidth]{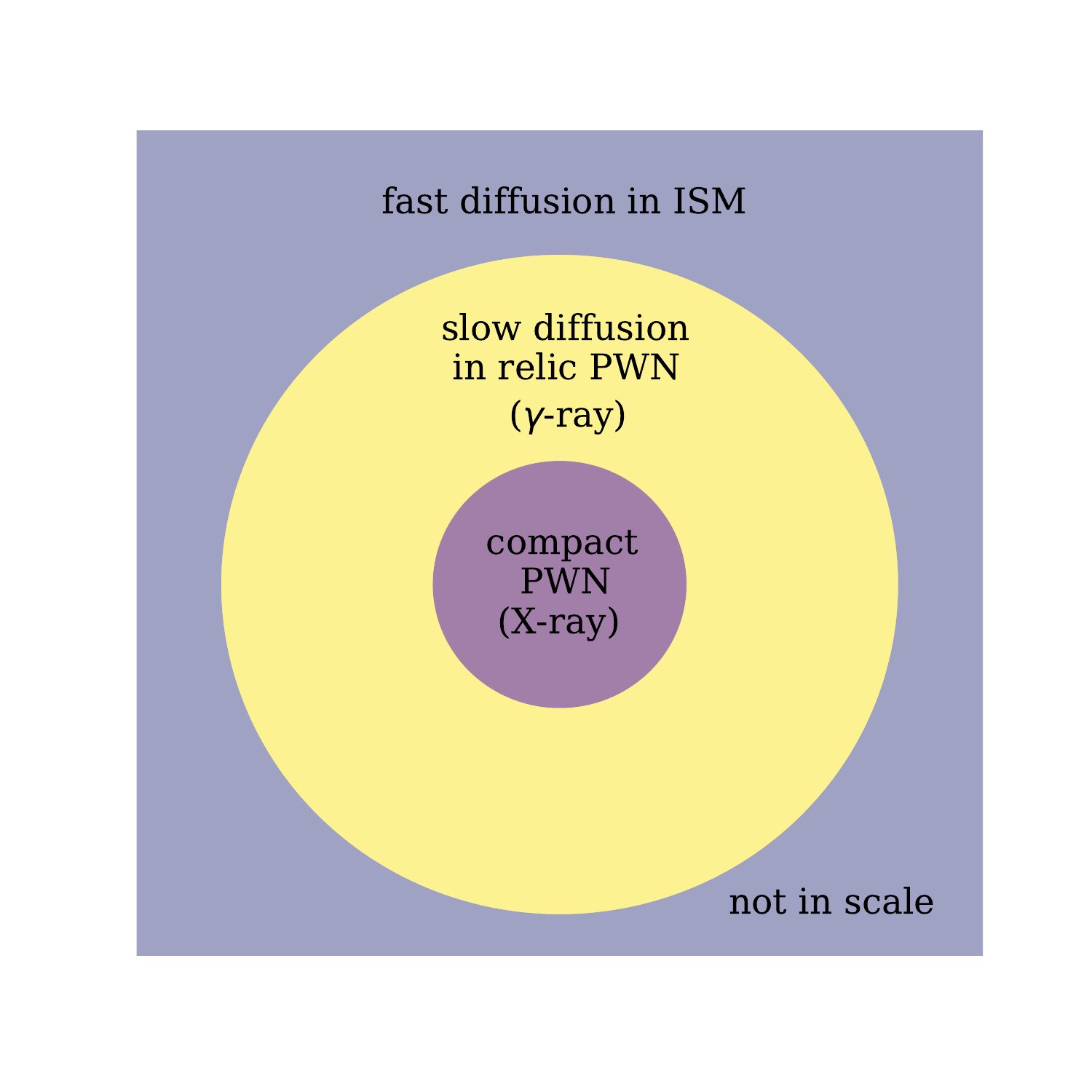} 
\caption{A schematic figure of particle transport in the Geminga PWN. The inner part of the PWN corresponds to the small bow shock nebula detected in X-ray and is unimportant in our model. The outer part of the PWN corresponds to the large nebula detected in $\gamma$-ray. Beyond the nebula, particles diffuse isotropycally in the ISM where the diffusion coefficeint is larger. } 
    \label{fig:schematic_3zone}
\end{center}
\end{figure}

We assume that the Geminga PWN has two parts as illustrated in Fig. \ref{fig:schematic_3zone}. The inner part is a compact nebula containing freshly injected particles. It corresponds to the bow shock nebula detected in X-ray with a size of $0.2{\rm pc}\,(d/250\rm pc)$ \citep{Posselt17}.  The outer part is the relic PWN dominated by aged particles that were injected a long time ago. The $\gamma$-ray emission region reported by {\it HAWC} and {\it MILAGRO} with a size of several tens of pc probably correspond to the relic PWN. Since the observed X-ray nebula is much smaller than the $\gamma$-ray nebula, we neglect it and focus on only the relic nebula and the surrounding ISM in the rest of the paper. 

The diffusion coefficient in the relic nebula is expected to be smaller than that in the ISM, as the magnetic field in the nebula is higher and the magnetic topology is also more complicated. This naturally explains why the $HAWC$ collaboration derived within the TeV emission region a diffusion coefficient much smaller than the ISM value. We propose that inefficient diffusion in the relic PWN of an old pulsar is likely to be a common phenomenon which can be tested by future $\gamma$-ray observation of other nearby pulsars. 

\section{$\gamma$-ray spectrum of Geminga PWN}{\label{sec:IC_emission}}
We turn now to calculate the IC emission of the Geminga PWN and compare it with the $\gamma$-ray data shown in Table \ref{table:PWN_flux}

\subsection{Energy losses}
When the electron-positron pairs diffuse in the PWN and the ISM, they lose energy to IC and synchrotron emission. The energy loss rate, including the Klein-Nishina effect, is
\begin{eqnarray}
\dot{E}&=&\dot{E}_{IC} +\dot{E}_{syn} \nonumber\\
&=& -\frac{4}{3}c\sigma_T \gamma^2\left[ \sum_i\frac{U_{i}}{(1+4\pi \gamma \epsilon_i)^{3/2}} + \frac{B^2}{8\pi}\right],
\label{eq:energy_loss}
\end{eqnarray}
where $c$ is the speed of light, $\gamma$ is the electron's Lorentz factor and $\sigma_T$ is the Thomson cross section. $U_i$ represents the energy density of the radiation field in different energy channel and $\epsilon_i$ is the corresponding normalized photon energy. For a Black Body radiation with a temperature $T$, we have $\epsilon=2.8K_BT/m_ec^2$, where $K_B$ is the Boltzmann constant and $m_e$ is the electron mass \citep{Moderski05}. In the calculation, we take into account CMB, infrared (IR), and optical photon fields, which are listed in Table \ref{table:photon_field}. $B$ is the magnetic field due to the central pulsar. 
The equipartition magnetic field in the X-ray tail close to the pulsar is estimated to be $20 \mu G$ \citep{Posselt17}, which implies that the spatial averaged $B$ field within the $\gamma$-ray nebula is likely marginally larger than the interstellar value of  a few $\mu G$. In this work, we adopt $B=3\mu G$ in the ISM and assume for simplicity that the cooling in the PWN and ISM is the same.

\begin{table}
\centering
\caption{Physical parameter of the photon fields}
\begin{threeparttable}
\begin{tabular}{ccc}
\hline\hline
photon field& $T$(K) &$U\rm (eV/cm^3) $  \\
\hline
CMB &2.7 &0.26 \\
IR &20 & 0.3 \\
Optical &5000 &0.3 \\
\hline\hline
\end{tabular} 
\end{threeparttable}
\label{table:photon_field}
\end{table}


The cooling time for an electron or positron to decay from $E_1$ to $E_2$ via synchrotron and IC emission is 
\begin{eqnarray}
\tilde{t}_c(E_1,E_2)&=&\int_{E_2}^{E_1} \frac{-dE}{\dot{E}}  \\
&=&\int_{E_2}^{E_1} \frac{3dE}{4\gamma^2\sigma_T c}\left[\frac{B^2}{8\pi}+\sum_i\frac{U_{i}}{(1+4\pi \gamma \epsilon_i)^{3/2}}\right]^{-1}. \nonumber 
\label{eq:cooling_time}
\end{eqnarray}

\subsection{IC emission}
Based on the conservation of total number of electrons and positrons, the accumulated differential number density of electrons or positrons at time $t_{age}$ satisfies
\begin{equation}
N(E) = \frac{1}{dE}\int_0^{t_{age}}Q(E_0,t_0)\, dE_0\, dt_0.
\end{equation}
A particle with energy $E$ now (i.e., at $t_{age}$) was originally injected at $t_0$ with energy $E_0$, i.e., $t_0 = t_{age}-\tilde{t}_c(E_0,E)$. The above integral can be further simplified into\begin{equation}
N(E) = -\int^{{\rm min}[E_{hi},\,E_0]}_E \frac{Q[E_0, t_{age}-\tilde{t}_c(E_0, E)] }{\dot{E}}\, dE_0,
\label{eq:SED}
\end{equation}
if we apply the relation $dt_0/dE = -d\tilde{t}_c(E_0,E)/dE=1/\dot{E}$. The upper limit of the integration ${\rm min}[E_{hi},\,E_0]$ accounts for the high energy cutoff of the injection spectrum.

The corresponding IC emisison from both electrons and positrons are calculated with the $naima$ python package \citep{Zabalza15}, which implements the analytical approximations to IC scattering of blackbody radiation developed by \cite{Khan14}. The formula remains accurate within one percent over a wide range of energies.

\begin{figure}
\begin{center}
\includegraphics[width=\columnwidth]{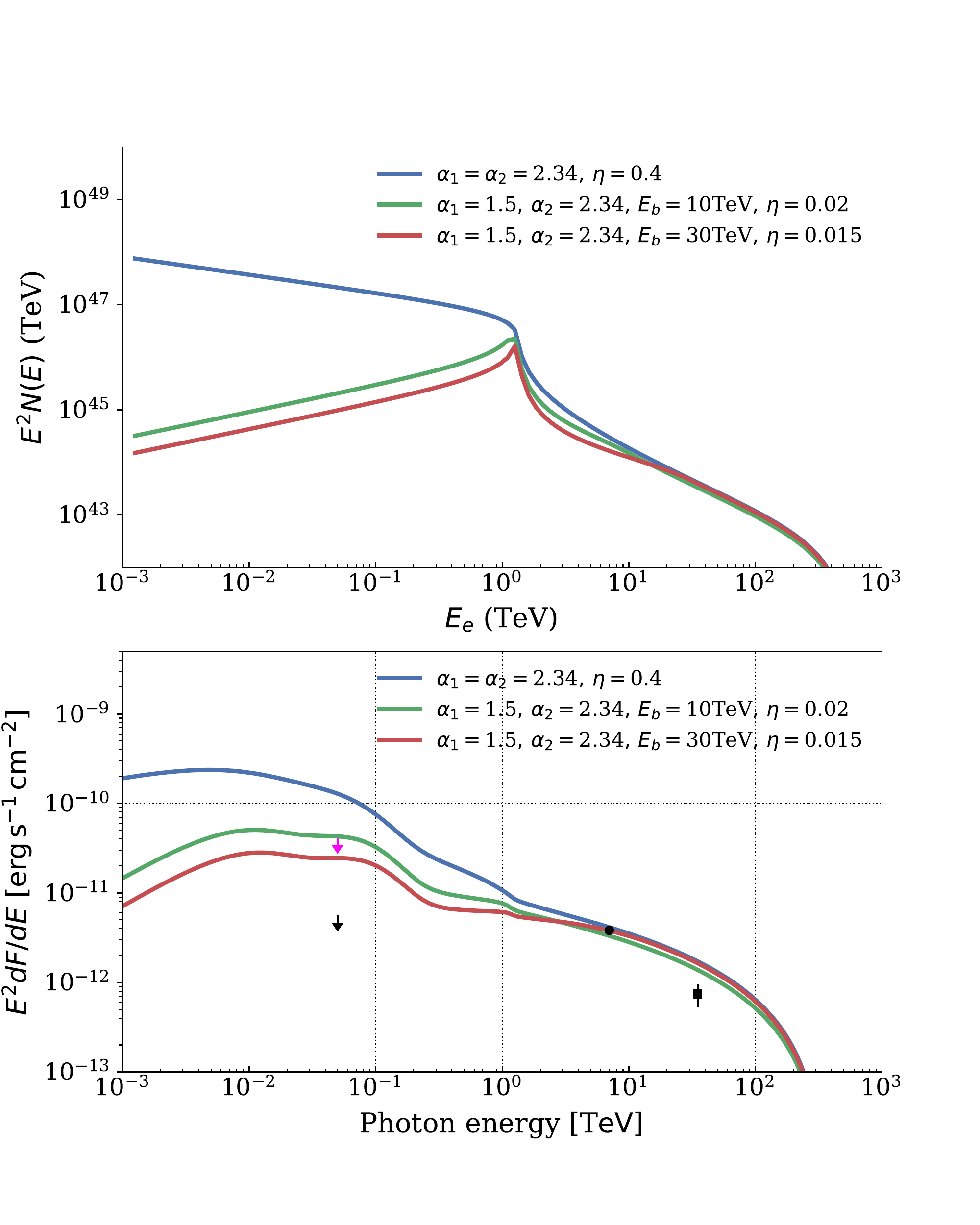} 
\caption{Upper panel: The differential number density $E^2N(E)$ as a function of the electron or positron energy for different $\alpha_1$ and $\alpha_2$. Lower panel: IC emission for different injection spectra. The black square and dot denote the $MILAGRO$ and $HAWC$ observations respectively. The magenta and black arrows describe the corrected and uncorrected $MAGIC$ upper limits respectively.} 
    \label{fig:IC_SED}
\end{center}
\end{figure}
\begin{figure}
\begin{center}
\includegraphics[width=\columnwidth]{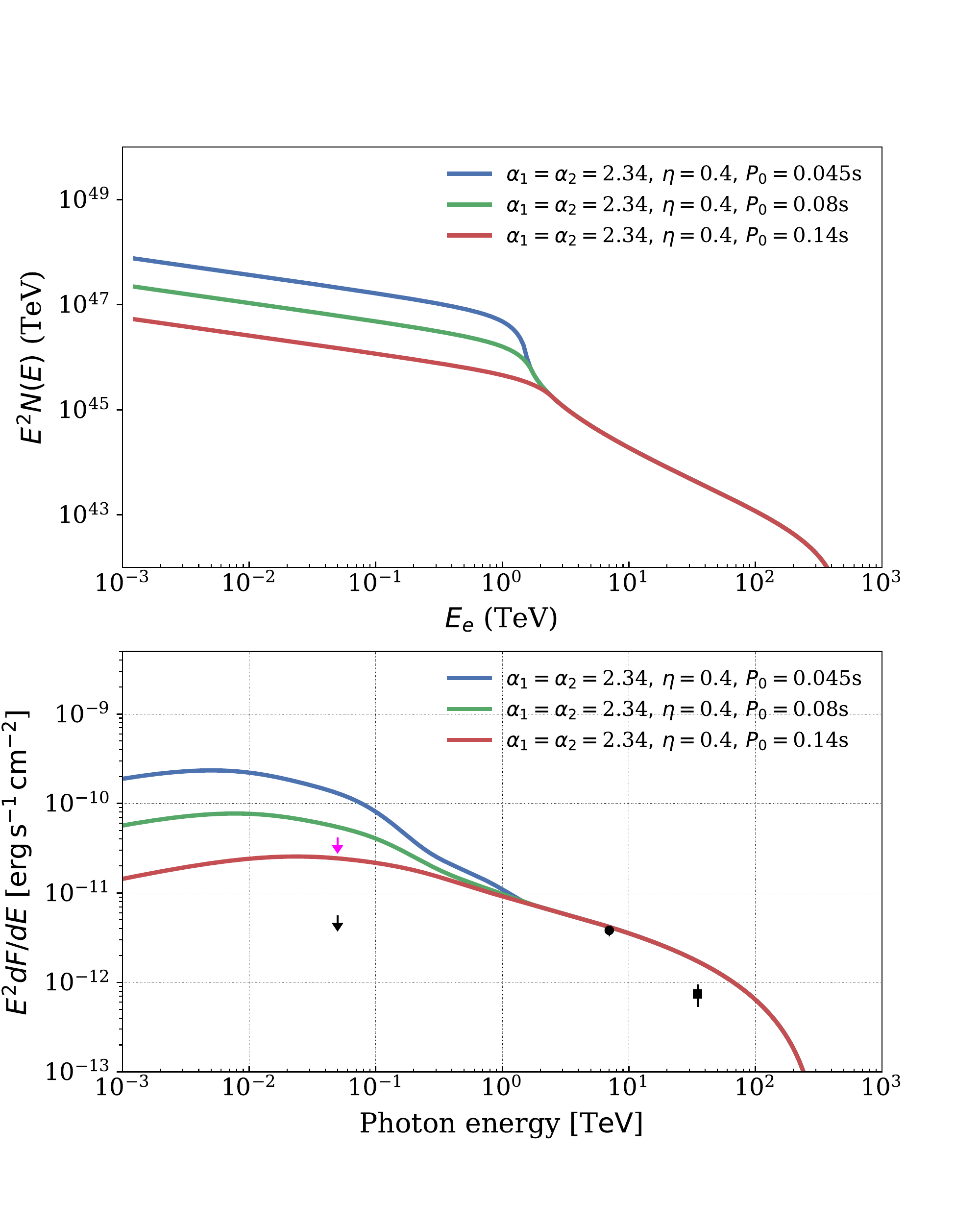} 
\caption{Same as Fig. \ref{fig:IC_SED} but for a single power law injection spectrum with different initial periods $P_0$. } 
    \label{fig:IC_SED_P0}
\end{center}
\end{figure}

In Fig. \ref{fig:IC_SED} and \ref{fig:IC_SED_P0}, we compare the IC emission of the nebula with the $\gamma$-ray data for different injection spectrum and different initial pulsar periods. The upper panel depicts the differential number density $E^2N(E)$ of the electrons or positrons, while the lower panel depicts the corresponding IC emission. The {\it MILAGRO}, {\it HAWC} and {\it MAGIC} data are illustrated with a black square, a circle and an arrow respectively. The single power law model developed in \cite{Abe17} overproduces the $\gamma$-ray emission in the GeV band even with the corrected $MAGIC$ upper limit. In order to explain the $MAGIC$ upper limit, we  need either a broken power law injection spectrum or a single power law spectrum but with a larger initial period $P_0$ as demonstrated in Fig. \ref{fig:IC_SED} and \ref{fig:IC_SED_P0} respectively. 

In the upper panel, the blue line shows a sharp drop around $1$TeV. This is mainly a reflection of the evolution history of the pulsar spin down luminosity as indicated in eq. (\ref{eq:luminosity_evolution}). Electrons or positrons with energy $E\gtrsim 1\rm TeV$ have a cooling time $\tilde{t}_c(E_{hi}, E)\lesssim t_{age}-\tau_0$ and are injected at $t_0\gtrsim \tau_0$ accordingly. $\tau_0$, the initial spin down timescale, is much smaller than $t_{age}$ with the default $P_0$. When $t_0\gtrsim \tau_0$, the pulsar spin down luminosity decreases rapidly with $t_0$. Hence, $N(E)$ steepens with energy due to the decrease in the pulsar luminosity. If we increase the initial period $P_0$, the sharp drop gradually disappears as shown in Fig. \ref{fig:IC_SED_P0}. This is because $\tau_{0}$ becomes comparable to $t_{age}$ and the pulsar spin down luminosity becomes for a short while a constant in time. $P_0\gtrsim 0.14$s is required to explain the {\it MAGIC} upper limit. 

According to Fig. \ref{fig:IC_SED} and \ref{fig:IC_SED_P0}, it is difficult to distinguish between the two explanations with only spectral information. The low energy surface brightness profile of the nebula is important to disentangle the two possibilities.


\section{Particle transport and spatial distribution}{\label{sec:spatial_distribution}}
We turn now to explore the spatial distribution of the electron-positron pairs and the positron flux on Earth with a two zone diffusion model. The two zone diffusion model was studied by \cite{Fang18} recently with numerical simulation. Here, we derive an analytical solution for the model, which involves a slow diffusion in the relic nebula and a fast diffusion in the ISM. The diffusion coefficient in the relic nebula is expected to be smaller than that in the ISM due to the larger magnetic field and the more complicated magnetic field topology\footnote{Note that for simplicity when calculating the cooling we take the magnetic field to be the same in the PWN and in the ISM}. The two zone diffusion model can explain the small diffusion coefficient derived by \cite{Abe17}, as it corresponds to the value in the relic nebula instead of the ISM. 


To obtain the spatial distribution of the pairs, we solve the particle transport equation
\begin{eqnarray}
\frac{\partial N(E, r, t)}{\partial t}&=&\nabla \cdot \left[D(E, r)\nabla N(E, r,t)\right] \nonumber\\
&&+\frac{\partial}{\partial E}\left[\dot{E}(E)N(E,r,t) \right]
 + Q(E,t )\delta(\vec{r}), \\ \nonumber
\label{eq:particle_transport}
\end{eqnarray}
where $D(E,r)$ is the diffusion coefficient. The source function $Q(E,t)$ and the energy loss term $\dot{E}(E)$ are defined in eq. (\ref{eq:source_function}) and (\ref{eq:energy_loss}) respectively. 
We begin with a discussion of homogeneous diffusion, i.e.,  a single zone diffusion, and then derive the solution for the two zone case.

\subsection{A single zone diffusion}
\begin{figure}
\begin{center}
\includegraphics[width=\columnwidth]{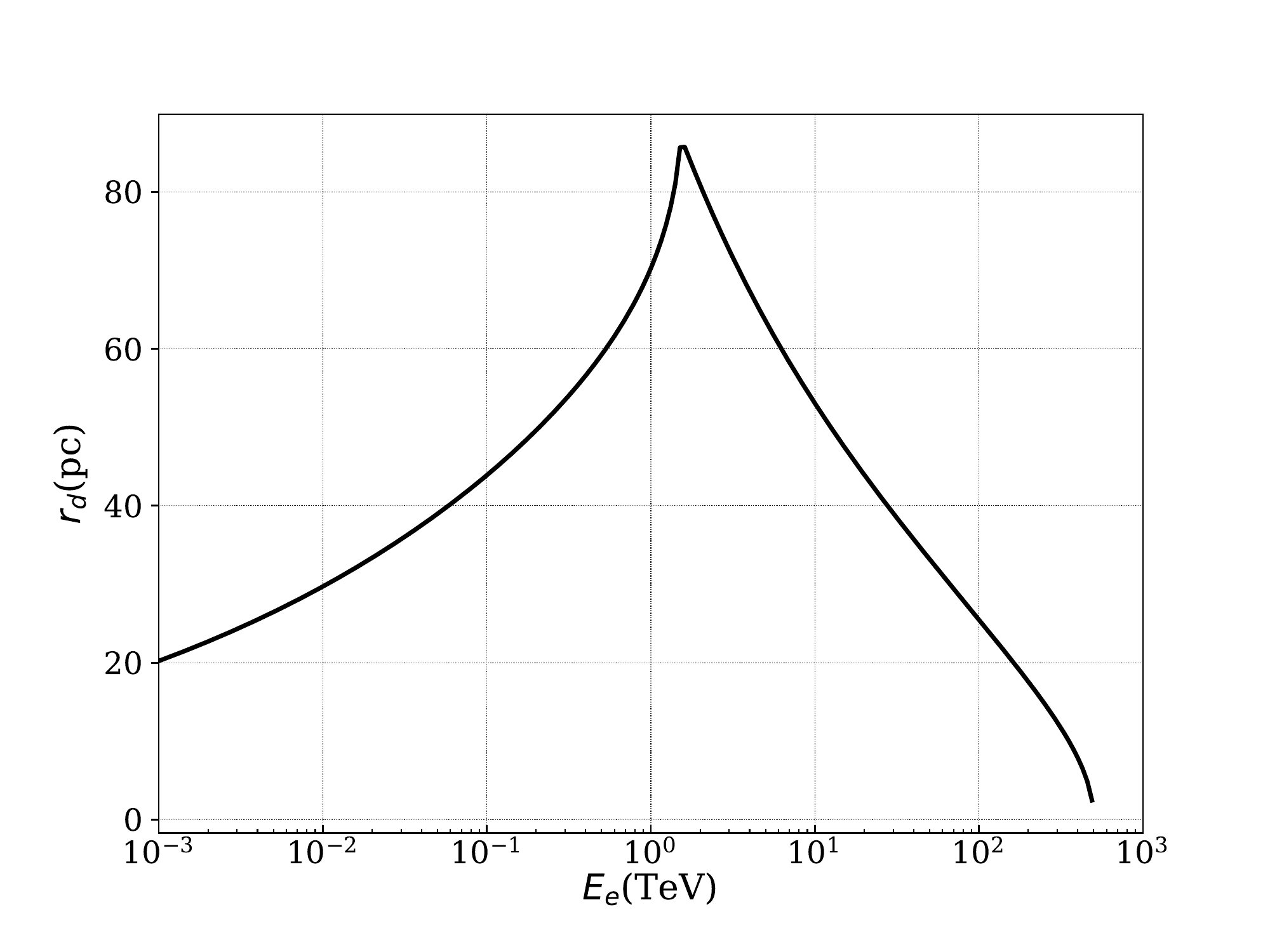} 
\caption{The diffusion length scale $r_d$, defined in eq. (\ref{eq:diffusion_length}), as a function of the electron or positron energy with homogeneous diffusion. } 
    \label{fig:diffusion_length}
\end{center}
\end{figure}

For a homogenous diffusion, \cite{Atoyan95} derive an analytical solution for the transport equation 
\begin{equation}
N_1(r, E) = \int_{{\rm max}[0,\,t_{age}-\tilde{t}_c(E_{hi}, E)]}^{t_{age}} \frac{\dot{E}(E_0)}{\dot{E}(E)}\frac{Q(E_0,t_0) }{\pi^{3/2}r_{d}^3}e^{-r^2/r_{d}^2} dt_0, 
\label{eq:solution_onezone}
\end{equation}
where $E_{hi}$ is the high energy cutoff of the injection spectrum and $\tilde{t}_c(E_{hi}, E)$ is the cooling time from $E_{hi}$ to $E$. If $E$ is close to $E_{hi}$, the cooling time $\tilde{t}_c(E_{hi},E)$ becomes smaller than $t_{age}$. In this situation, only particles injected at $t_0>t_{age}-\tilde{t}_c(E_{hi}, E)$ contribute to $N_1(r, E)$. This is reflected in the lower limit of the integration. $E_0$ corresponds to the initial energy of an electron or positron at injection time $t_0$ that cooled down to $E$ at $t_{age}$. $r_d$ is the diffusion length scale which is defined as
\begin{equation}
r_{d} = 2 \left[\int_E^{min[E_0,E_{hi}]}\frac{D(\hat{E})}{\dot{E}(\hat{E})} d\hat{E}\right]^{1/2}.
\label{eq:diffusion_length}
\end{equation}
In Fig. \ref{fig:diffusion_length}, we plot $r_{d}$ as a function of $E_e$ for $t_{age}=320\rm kyr$, $B=3\mu$G and
\begin{equation}
D=4.5\times 10^{27}\rm cm^2/s \left(\frac{E}{100TeV}\right)^{1/3}
\label{eq:diffusion_coefficent_onezone}
\end{equation}
for a Kolmogorov type turbulence.

$\dot{E}(E_0)/\dot{E}(E)$ in eq. (\ref{eq:solution_onezone}) characterizes the energy loss of a particle for an arbitrary cooling function. If there are no energy losses or very slow cooling, then $\dot{E}(E_0)/\dot{E}(E) \rightarrow 1$ and eq. (\ref{eq:solution_onezone}) recovers the solution with no energy loss, i.e., 
\begin{equation}
N_{1,nl}(r, E) = \int_{0}^{t_{age}} \frac{Q(E,t) }{\pi^{3/2}r_{d}^3}e^{-r^2/r_{d}^2} dt.
\end{equation}

In eq. (\ref{eq:solution_onezone}), the term
\begin{equation}
H_1(r,E)=\frac{1}{\pi^{3/2}r_{d}^3}e^{-r^2/r_{d}^2}
\end{equation}
describes the spatial distribution of all the particle. If integrated over the whole space, eq. (\ref{eq:solution_onezone}) simply recovers the spectral energy distribution derived in eq. (\ref{eq:SED}).

\subsection{Two zone diffusion}{\label{sec:twozone_diffusion}}
We assume now a diffusion coefficient of the form:
\begin{equation}
D(E, r)=\left(\frac{E}{100 \rm TeV}\right)^{1/3}\begin{cases}
D_1, \, 0<r<r_{b},\\
D_2, \, r\geq r_{b},
\end{cases}
\label{eq:diffusion_coefficient}
\end{equation}
where $r_b$ is the boundary between the two zones. For simplicity we fix $r_b=60$pc and then vary $D_1$ to reproduce the HAWC surface brightness profile \citep{Abe17}. After obtaining $D_1$, we vary $D_2$ and compare the results with the positron excess data. $r_b$ is expected to be  $\gtrsim 30$pc, otherwise we can't reproduce the $HAWC$ surface brightness profile. If $r_b$ is too large, then the positron flux on Earth becomes similar to the single zone model result and it cannot account for the positron excess. The exact value of $r_b$ can be constrained by future multi-wavelength observation.

To derive the solution for two zone diffusion, we only need to replace the homogeneous diffusion term $H_1(r, E)$ in eq. (\ref{eq:solution_onezone}) with the two zone diffusion term $H_2(r,E)$. The solution now becomes
\begin{equation}
N_2(r, E) = \int_{{\rm max}[0,\,t_{age}-\tilde{t}_c(E_{hi}, E)]}^{t_{age}} \frac{\dot{E}(E_0)}{\dot{E}(E)}Q(E_0,t_0) H_2(r,E) dt_0, 
\label{eq:solution_twozone}
\end{equation}
where
\begin{equation*}
H_{2}(r,E) = \frac{ b(b+1)}{\pi^{3/2}r_{d1}^3[\rm 2b^2erf(r_b)-b(b-1)erf(2r_b)+2erfc(r_b)]}
\end{equation*}
\begin{eqnarray}
\times \begin{cases}
\left[e^{-r^2/r_{d1}^2} + \left(\frac{b-1}{b+1}\right)\left(\frac{2r_b}{r}-1\right)e^{-(r-2r_{b})^2/r_{d1}^2} \right], \, r<r_b,\\
\left(\frac{2b}{b+1}\right)\left[\frac{r_b}{r}+b(1-\frac{r_b}{r})\right]e^{-[(r-r_{b})/r_{d2}+r_{b}/r_{d1}]^2}, \, r\geq r_b.\\
\end{cases}&&
\label{eq:two_zone_diffusion}
\end{eqnarray}
$b$, $r_{d1}$  and $r_{d2}$ are constants and are defined as 
\begin{equation}
b\equiv \frac{\sqrt{D_1}}{\sqrt{D_2}},
\end{equation} 
\begin{equation}
r_{d1}(E) \equiv 2 \left[\int_E^{min(E_0,E_{hi})}\frac{D(x,r<r_b)}{\dot{E}(x)} dx\right]^{1/2}
\end{equation}
and
\begin{equation}
r_{d2}(E) \equiv 2 \left[\int_E^{min(E_0,E_{hi})}\frac{D(x,r>r_b)}{\dot{E}(x)} dx\right]^{1/2}
\end{equation}
respectively.
erf($x$) is the error function and  erfc($x$)=1-erf($x$).
Appendix \ref{sec:twozone_solution} describes a detailed derivation of $H_2(r,E)$.

\begin{figure}
\begin{center}
\includegraphics[width=\columnwidth]{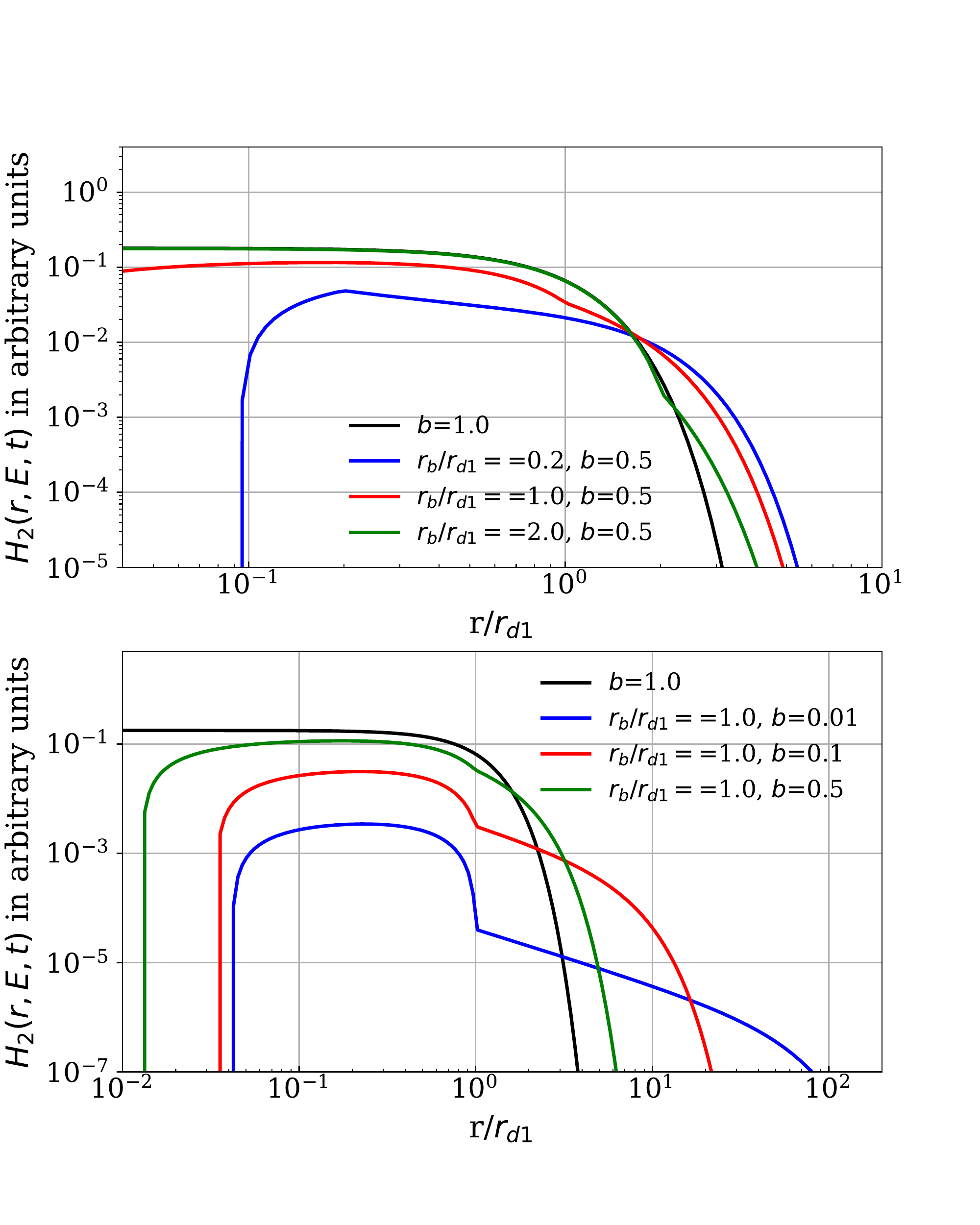} 
\caption{The two zone diffusion term $H_2(r,E,t)$ defined in eq. (\ref{eq:two_zone_diffusion}) as a function of the dimensionless length scale $r/r_{d1}$ for different combinations of $b$ and $r_b/r_{d1}$. The black solid line is the result from a single zone diffusion with $b=1$. } 
    \label{fig:two_zone_diffusion}
\end{center}
\end{figure}

$r_{d1}^3 H_2(r,E)$ is a dimensionless quantity, which depends on three dimensionless parameter $r/r_{d1}$, $b$ and $r_b/r_{d1}$. Here, we are interested in the regime with $b<1$ and $r_b/r_{d1}\gtrsim 1$. Since the diffusion coefficient in the ISM is expected to be larger than that in the nebula. Under the assumption of $r_b=60$pc, the diffusion length scale presented in Fig. \ref{fig:diffusion_length} satisfies $r_{d1}\lesssim r_b$ for the entire energy range. 

In the upper panel of Fig. \ref{fig:two_zone_diffusion}, we fix $b$ and then investigate how does the spatial distribution of $H_2$ vary with $r_b/r_{d1}$. As $r_b/r_{d1}$ increases the resulting profile gradually approaches the single zone solution. This is mainly because when $r_b\gg r_{d1}$, the source is  unable to feel the second zone. In the lower panel, we fix $r_b/r_{d1}$ and study how does $H_2$ vary with $b$. For a given ratio of $r_b/r_{d1}$, when $b$ decreases, particles can diffuse further away from the source. This helps to explain the observed positron excess. The effect is mainly important for the particles with $r_{d1}\approx r_b=60$pc, i.e., particles with energy around a few TeV (see Fig. \ref{fig:diffusion_length}). 
$H_2$ approaches zero at small $r/r_{d1}$ values because we adopt an absorption boundary at the source. It is not clear what should be the appropriate condition at the inner boundary. But we expect that the inner boundary condition has only small effects on the large scale structure like the positron flux on Earth.

\subsection{The surface brightness}
\begin{figure}
\begin{center}
\includegraphics[width=\columnwidth]{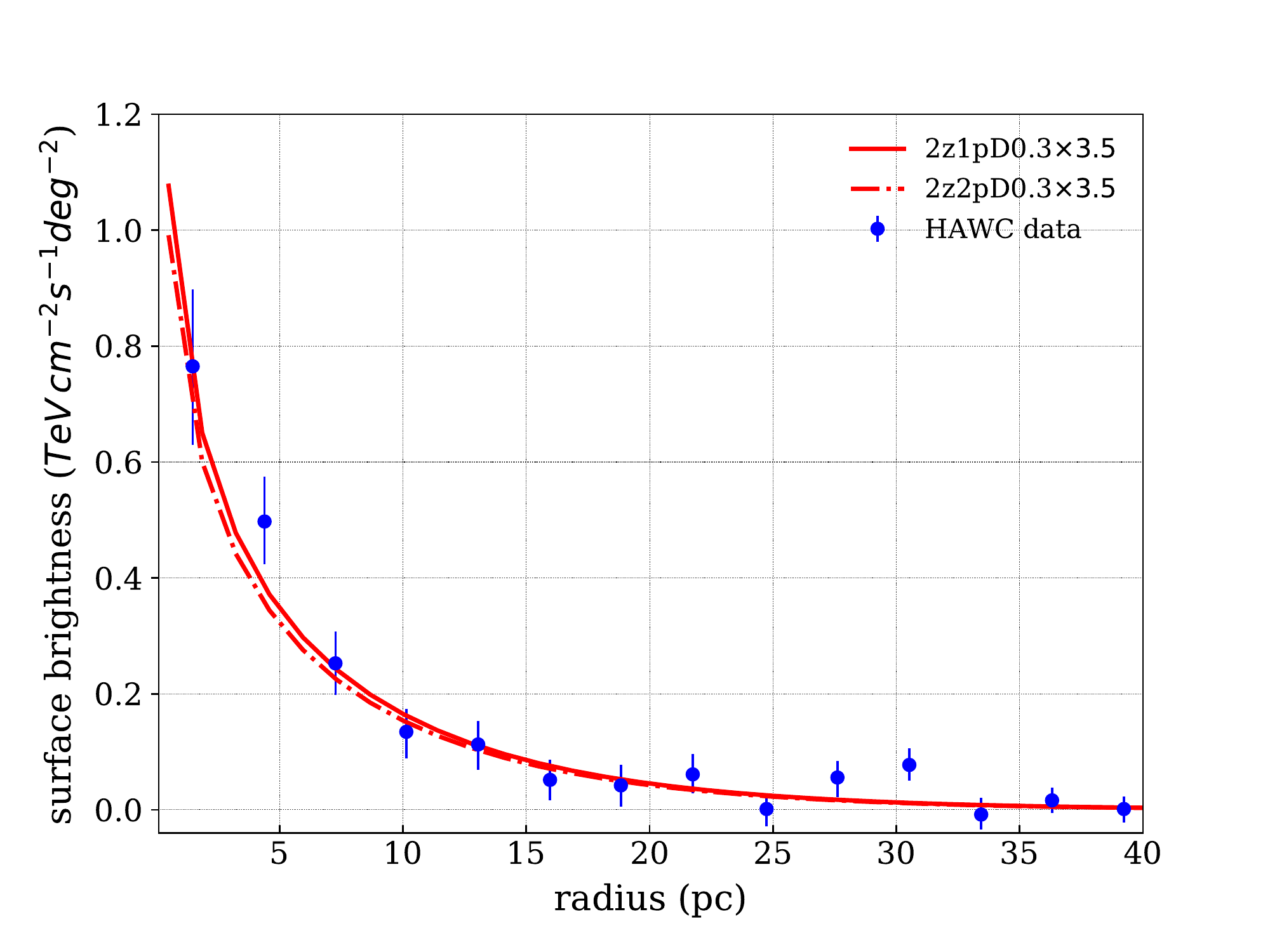} 
\caption{Surface brightness profiles between $5$ and $50$TeV of the photon energy. The {\it HAWC} data is taken from \citet{Abe17} and the {\it HAWC} model is calculated with the same parameter as in \citet{Abe17}. 2z1pD03 and 2z2pD03 are the single power law and broken power law model respectively.} 
    \label{fig:surface_brightness}
\end{center}
\end{figure}

Integrating $N_2(r,E)$ along the line of sight and then calculating the corresponding IC emission, we obtain the $\gamma$-ray surface brightness profile of the Geminga PWN (see Fig. \ref{fig:surface_brightness}). We neglect the finite size of the nebula, which may introduce at most an effect of $(r_{d}/d)^2< 10\%$.  To simplify the calculation, we also approximate 
\begin{equation}
r_{d}(E)\approx 2\left[\frac{D(E) E}{(1-\delta)\dot{E}}\right]^{1/2}\left[1 - \left(\frac{E}{E_0}\right)^{1-\delta}\right]^{1/2}.
\end{equation}
and
\begin{equation}
E_0(E, \tilde{t}_c )\approx \frac{E}{1-\dot{E} \tilde{t}_c /E},
\end{equation}
where $\tilde{t}_c $ is the cooling time since injection.

In Fig. \ref{fig:surface_brightness}, we compare the {\it HAWC} data with different model results. The physical paramters applied in the different models are listed in Table \ref{table:model_parameter}. The "HAWC" model is calculated with the same parameter as in \cite{Abe17}. "2z1pDx" stands for a two zone diffusion model with a single power law injection apectrum and $D_1/D_2=x$, while  "2z2pDx" denotes two zone diffusion model with a broken power law injection spectrum and $D_1/D_2=x$. 
Since we adopt $r_b=60$pc, which is larger than the size of observed $\gamma$-ray nebula, the two zone diffusion model produces almost the same surface brightness profile as the single zone model within $r_b$. 
Within the two zone model the surface brightness profile at radius $r\lesssim 20$pc is insensitive to the $D_1/D_2$ ratio. Hence, models with $D_1/D_2= 0.05$ and $0.01$  are not plotted in Fig. \ref{fig:surface_brightness}.

We multiply our results by a factor of 3.5 when comparing the $HAWC$ data. It is because when we apply the same parameter as in \cite{Abe17}, we found that the resulting $\gamma$-ray flux is consistent with the value based on the disk template\footnote{The $HAWC$ data shown in Fig.  \ref{fig:IC_SED} and \ref{fig:IC_SED_P0} are based on the disk template.}, see the supplementary Fig. 2 in \cite{Abe17}. The surface brightness data however appears to agree with the value estimated with a diffusion template, which is larger than the disk template's value by a factor about 3.5. This discrepancy won't affect the discussion with broken power law injection spectrum, as we can simply increases the acceleration efficiency by the same factor. Larger $\gamma$-ray flux of the Geminga PWN does make it difficult for the model with a single power law injection spectrum to interpret the surface brightness data as the current $\eta=0.4$.

\begin{table*}
\centering
\caption{Model parameter}
\begin{threeparttable}
\begin{tabular}{lccccccccccc}
\hline\hline
Model name& $\alpha_1$ &$\alpha_2$ & $E_b$(TeV)  &$D_1$($10^{27}\rm cm^2/s$) & $D_1/D_2$&$r_b$(pc) & $P_0$ (s) & $\eta$ & $B$ ($\mu$G)& $E_{min}$(TeV) &$E_{max}$(TeV)\\
\hline
HAWC &2.34 &2.34 & -- &4.5 & 1 & -- & 0.045 & 0.4 & 3&$10^{-3}$ & $500$\\
2z1pD03 &2.34 &2.34 & -- &4.5 & 0.3 & 60 & 0.14 &  0.4 & 3& $10^{-3}$ & $500$\\
2z1pD005 &2.34 &2.34 & -- &4.5 & 0.05 & 60 & 0.14 &  0.4 & 3&$10^{-3}$ & $500$\\
2z1pD001 &2.34 &2.34 & -- &4.5 & 0.01 & 60 & 0.14 &  0.4 & 3& $10^{-3}$ & $500$\\
2z2pD03 &1.5 & 2.34 & 30 & 4.5 & 0.3 & 60 &0.045 & 0.015 & 3&$10^{-3}$ & $500$\\
2z2pD005 &1.5 & 2.34 & 30 & 4.5 & 0.05 & 60 &0.045 & 0.015 & 3&$10^{-3}$ & $500$\\
2z2pD001 &1.5 & 2.34 & 30 & 4.5 & 0.01 & 60 &0.045 & 0.015 & 3&$10^{-3}$ & $500$\\
\hline\hline
\end{tabular} 
$D_1$ is at 100TeV 
\end{threeparttable}
\label{table:model_parameter}
\end{table*}

\subsection{The positron flux at Earth}
\begin{figure}
\begin{center}
\includegraphics[width=\columnwidth]{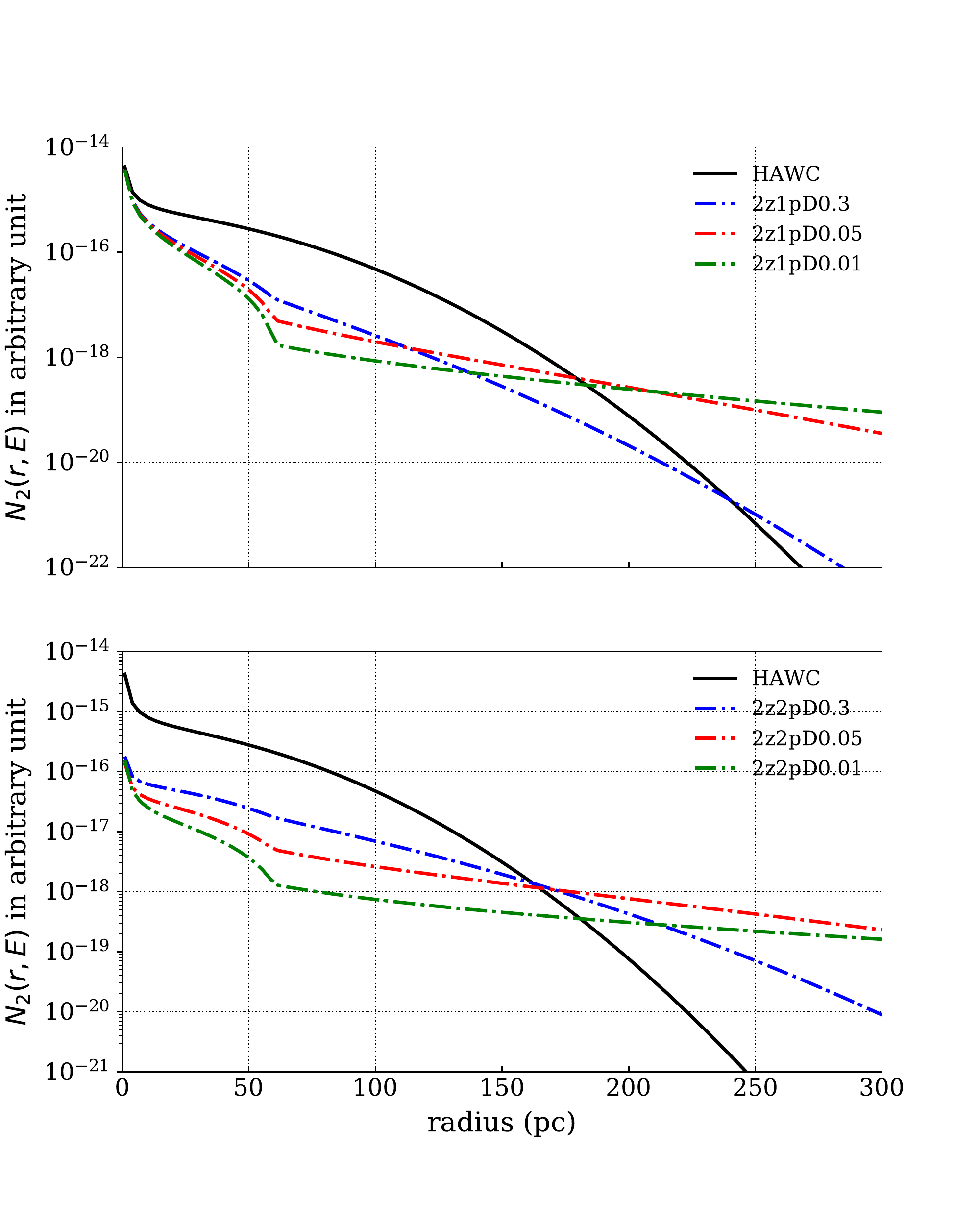} 
\caption{The differential number density of positron (or electron), $N_2(r,E)$, as a function of distance $r$ at energy $E=1\,$TeV. A single power law with a larger initial period $P_0$ (upper panel). A broken power law injection spectrum with a smaller $P_0$ (lower panel). The black solid line represents the model from \citet{Abe17}.} 
    \label{fig:twozone_number_density}
\end{center}
\end{figure}

Fig. \ref{fig:twozone_number_density} depicts the number density of positrons, $N_2(r,E)$, as a function of the radius $r$ at energy $E=1$TeV. We focus on two different situations, a single power law injection spectrum with a larger initial period $P_0$ and a broken power law spectrum with a smaller $P_0$. This is illustrated in the upper panel and the lower panel respectively. The basic parameters for all the different models are listed in Table \ref{table:model_parameter}. 

When the $D_1/D_2$ ratio decreases, the positron flux at Earth (assuming $d=250$pc) increases as the Earth position is shifted from the exponential tail to the plateau region in the spatial distribution (see Fig. \ref{fig:twozone_number_density}). The positron flux reaches a maximum at $D_1/D_2 \sim 0.05$ and it is boosted by more than 2 orders of magnitude compared with the single zone model results. When the $D_1/D_2$ ratio decreases further, the positron flux at Earth decreases as the normalization constant of the plateau region becomes smaller. 

\begin{figure}
\begin{center}
\includegraphics[width=\columnwidth]{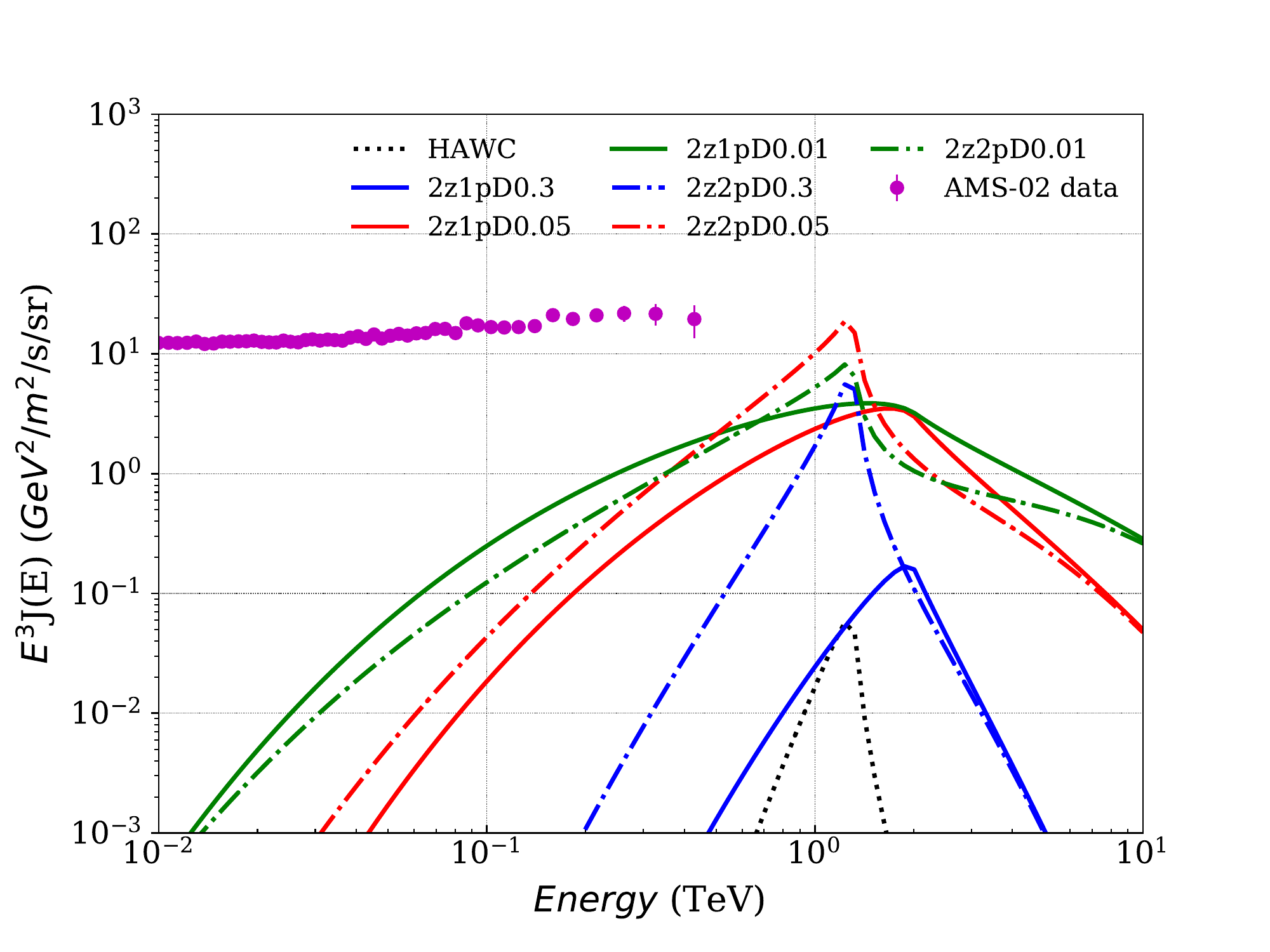} 
\caption{The positron flux at Earth produced by the Geminga pulsar with different model setup listed in Table \ref{table:model_parameter}.  } 
    \label{fig:positron_excess}
\end{center}
\end{figure}

Fig. \ref{fig:positron_excess} depicts the positron spectrum produced by the Geminga pulsar at Earth for the different models shown in Fig. \ref{fig:twozone_number_density}. 
The single power law model produces a flat positron spectrum at Earth, which seems to be more consistent with data\footnote{Note that in the spiral arm model the pulsars contribute only to the higher end of the excess.}. However, the broken power law model cannot be ruled out. If $D_1/D_2\sim 0.05$, both the single power law model and the broken power law model can contribute a significant fraction of the positron excess above $\sim 300$GeV. It worth noting that all the models discussed here are not meant to be the best fit for the positron flux data but mainly for illustration. As it is difficult to explore the whole parameter space. Future $\gamma$-ray observations in the GeV band will enable us to put strong constraints on the model set up and narrow the parameter space. Positron flux from other nearby pulsars also need to be investigated in future study.

With $D_1/D_2=0.05$, the diffusion coefficient in the ISM is found to be
\begin{equation}
D_{ISM}\sim 2\times 10^{27}\rm cm^2/s \left(\frac{E}{1GeV}\right)^{1/3}.
\end{equation}
which is still one order of magnitude smaller than the standard value but more consistent with the low value required in the spiral arm model \citep{Ben14}. For a given average path length, CRs experience a lower average interaction with the ISM in the spiral arm model than in the standard CR model. In order to recover the observed secondary to primary ratio, the spiral arm model requires a smaller halo of about a few hundred pc to keep the CRs closer to the galactic plane where the density is higher. The smaller halo requires a lower diffusion coefficient in the ISM which is estimated to be $\sim 10^{27} \rm cm^2/s$ at 1GeV \citep{Shaviv09}. The two zone diffusion model studied here seems to support the spiral arm model.


\subsection{The proper motion of the Geminga pulsar}{\label{sec:proper_motion}}
The Geminga pulsar has a transverse velocity of $ v_t \approx 211 (d/250\rm pc) \rm \, km s^{-1}$ \citep{Faherty07}. 
The corresponding transverse distance traveled by the pulsar in $t_{age}$ is estimated to be 
\begin{equation}
d_{t}=v_t t_{age}\sim 70\, {\rm pc}\, \left(\frac{d}{250 {\rm pc}}\right)\left(\frac{t_{age}}{ 320\rm kyr}\right).
\label{eq:traverse_distance}
\end{equation} 
The displacement $d_t$ introduced by the pulsar proper motion is comparable to the spatial extension of the TeV nebula detected by HAWC \citep{Abe17}. This may affect the $\gamma$-ray morphology of the PWN. The radial velocity of the Geminga pulsar is unclear but it is likely smaller than its transverse velocity, otherwise the bow shock nebula won't be revealed clearly in X-ray due to projection effect \citep{Posselt17}. 

As a first attempt to investigate the effect of the proper motion, we focus on the simple single zone diffusion model. The two zone diffusion model requires knowledge of the time and spatial evolution of the PWN, which is beyond the scope of this work. According to section \ref{sec:observation_pulsar}, we choose that the pulsar is at the origin $O(0, 0, 0)$ now and it was moving along the x-axis with a transverse velocity of $v_t \approx 211\rm km/s$. The y-axis is perpendicular to the proper motion in the plane of sky, while the z-axis is along the line of sight. The birth place of the pulsar then is simply at ($-d_t, 0, 0$), where $d_t=v_t t_{age}\approx 70$pc. The differential number density in eq. (\ref{eq:solution_onezone}) is modified into
\begin{eqnarray}
N_1(x,y,z,E) &=& \int_{{\rm max}[0,\,t_{age}-\tilde{t}_c(E_{hi}, E)]}^{t_{age}} \frac{\dot{E}(E_0)}{\dot{E}(E)}\frac{Q(E_0,t_0) }{\pi^{3/2}r_{d}^3} \nonumber\\
&\times & e^{-[(x+v_{t}t_{age}-v_{t}t_0)^2+y^2+z^2]/r_{d}^2}\, dt_0,
\end{eqnarray}
where $v_t$ is the traverse velocity of the pulsar.

\begin{figure}
\begin{center}
\includegraphics[width=\columnwidth]{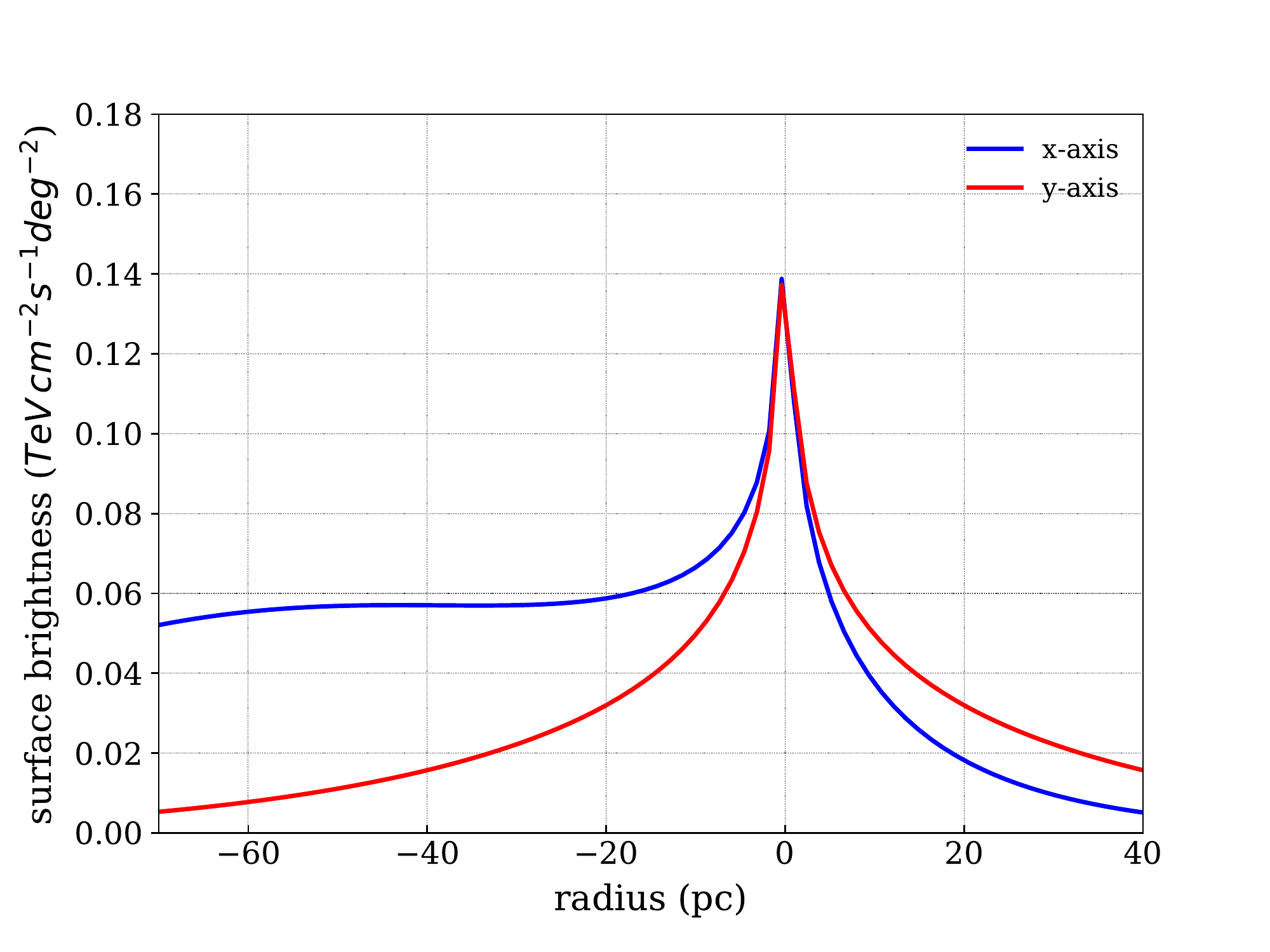} 
\caption{The surface brightness profile between $50$ and $100\,$GeV (i.e., the $MAGIC$ band) both along the proper motion direction (x-axis) and perpendicular to the proper motion direction in the plane of sky (y-axis).} 
    \label{fig:propertion_motion_SB}
\end{center}
\end{figure}

The proper motion of the Geminga pulsar is mainly important for the surface brightness profile of the PWN at low energies $E_e\lesssim 1$TeV. In the TeV band, the morphology of the nebula is found to be not strongly affected by the proper motion. Because the corresponding electron-positron pairs have a smaller life time and the corresponding transverse motion of the pulsar is smaller than that estimated in eq. (\ref{eq:traverse_distance}). It is consistent with the roughly spherical symmetric nebula revealed in the {\it HAWC} observations \citep{Abe17}. In the lower energy band, where the life time of electrons-positron pairs are close to $t_{age}$, a bow shock nebula morphology is instead expected. In Fig \ref{fig:propertion_motion_SB}, we present the surface brightness profile of the Geminga PWN between $50$ and $100$GeV that is relevant for the {\it MAGIC} band with the set up of the HAWC model. The nebula is elongated in the negative x direction due to the proper motion.  Within the single zone model, the proper motion of the pulsar has a limited influence on the positron flux at Earth, as it only changes the distance of the pulsar slightly introducing an effect $\lesssim (v_tt_{age}/d)^2 \approx 10\%$.

\section{Discussion} {\label{sec:discussion}}
We propose that the $\gamma$-ray emission detected by $HAWC$ and $MILAGRO$ in the direction of Geminga originated from the relic PWN surrounding the pulsar. We developed a two zone diffusion model with a slow diffusion $D_1$ in the PWN and a fast diffusion $D_2$ in the ISM.  This model would explain both the surface brightness profile and the positron excess $\gtrsim 300$GeV. With $D_1/D_2 \sim 0.05$, the Geminga pulsar can supply a significant fraction of the positron excess above $\sim 300$GeV. The diffusion coefficient in the ISM is $D_{ISM}\sim 2\times 10^{27}\rm cm^2/s (E/1GeV)^{1/3}$. This value is one order of magnitude smaller than the standard value but more consistent with the low value required in the spiral arm model for the CR propagation \citep[e.g.,][]{Shaviv09}.

There are several factors that can affect the $\gamma$-ray emission and the surface brightness profile of the Geminga PWN and the positron flux at Earth. The {\it MAGIC} upper limit implies either a broken power law injection spectrum or a single power law but with a larger initial period $P_0$. This can be tested by future multi-wavelength observation. We assume that the particle acceleration efficiency $\eta $ is a constant. If instead $\eta$ gradually decreases with time, then our calculation underestimates the positron flux from the Geminga pulsar. This helps to relieve the discrepancy between the HAWC detection \citep{Abe17} and the idea that the Geminga pulsar is a main source of the positron excess. The proper motion of the Geminga pulsar was studied with a single zone model. We found that a bow shock nebula morphology likely has appeared in the GeV emission. This can be tested by future $MAGIC$ observation. 

The main caveat of our model is that we neglect the dynamical evolution of the PWN and focus only on the time evolution of the pulsar spin-down power. In other words, we investigate a model with a dynamical pulsar and a static nebula with time independent diffusion coefficient and magnetic field. The peak of the positron flux on Earth corresponds to the pairs injected at an early phase of the Geminga pulsar, when the pulsar spin down luminosity was still large. Therefore, the understanding of the diffusion of TeV pairs in young PWNe, like the Crab, is crucial for explaining the observed positron excess at $\gtrsim 300$GeV. The slow diffusion revealed by \cite{Abe17} in the $\gamma$-ray emission region instead characterizes the properties of the relic nebula at the late phase and is likely irrelevant.  

In young PWN like the Crab with an age of a few thousand years, the diffusion coefficient of TeV positrons is found to be $\sim 5\times 10^{26}\rm cm^{2}/s$ through a spectral index fitting of the synchrotron emission\footnote{Note we extend the diffusion coefficient provided in Table 3 of \cite{TC12} to 1TeV with a Komogorov type energy dependence.} \citep{TC12}. The corresponding escape time is estimated to be
\begin{equation}
t_{esc}\sim \frac{R^2}{4D_{PWN}} \sim 150 \rm yr\,\left(\frac{R}{1\rm pc}\right)^2 \left(\frac{5\times 10^{26} \rm cm^2/s}{D_{PWN}}\right).
\end{equation}
In the early phase of the pulsar evolution, the nebula is small and $t_{esc}$ is much smaller than the age of the Geminga pulsar $t_{age}$. The diffusion time of TeV positrons in the ISM is approximately the age of the pulsar, i.e., $t_{age}=d^2/4D_{ISM}$. This implies 
\begin{equation}
D_{ISM} \sim \frac{d^2}{4t_{age}}\sim 1.5\times 10^{28}{\rm cm^2/s}\,\left(\frac{d}{250 \rm pc}\right)^2 \left(\frac{320 \rm kyr}{t_{age}}\right)
\end{equation}
at 1TeV, where $d$ is the distance of the pulsar. If we assume a Komogorov type turbulence, then the  diffusion coefficient in the ISM becomes $1.5\times 10^{27}\rm cm^2/s (E/1GeV)^{1/3}$ which again is more consistent with the value required by the spiral arm model. Based on the above discussion, the time evolution of PWNe can affect the positron flux on Earth. This will be addressed in future work.

In summary, by considering a physically motivated PWN model for the $\gamma$-ray observation of $HAWC$ and $MILAGRO$ we have shown that the Geminga pulsar is a good candidate to be source of a significant fraction of the observed positron excess at $\gtrsim 300$GeV .
\section*{Acknowledgments}
We thank Nir Shaviv and Reetanjali Moharana for helpful discussions. The research was supported by an ERC advanced grant TReX  and by the Che-Isf I-core center for excellence.

\appendix
\section{The particle distribution for two zone diffusion}{\label{sec:twozone_solution}}
In this Appendix, we derive the particle distribution for a two zone diffusion model using the solution of single zone diffusion. We denote by subscript 1 a single zone diffusion and subscript 2 a two zone diffusion.

The particle transport equation for one dimensional (1D) diffusion is
\begin{equation}
\frac{\partial N(x,t)}{\partial t}=\frac{\partial }{\partial x}\left(D(x)\frac{\partial N(x,t)}{\partial x}\right) +Q_0(E)\delta(x,t),
\label{app:eq:1Ddiffusion}
\end{equation}
where $N(x,t)$ is the line density of particles at a position $x$ and time $t$, $D(x)$ is the diffusion coefficient and $Q_0(E)$ is the injection constant. The particles are injected at $x=0$ and $t=0$ as indicated by the delta function $\delta(x,t)$.  We consider a two zone diffusion with
\begin{equation}
D(x)=\begin{cases}
D_1, x<x_{b},\\
D_2, x\geq x_{b},
\end{cases}
\end{equation}
where $D_1$ and $D_2$ are the diffusion coefficients for the two zones respectively. The solution of eq. (\ref{app:eq:1Ddiffusion}) depends on the relative position of the source and the contact discontinuity $x_b$. In the following discussion, we focus on the situation with $x_b>0$, while the solution for $x_b <0$ case can be obtained with the transformation $x,x_b \rightarrow -x,-x_b$.

In the single zone diffusion case (i.e., $D_1=D_2$), based on dimensional analysis the solution should satisfy
\begin{equation}
N_1(x,t) = \frac{Q_0}{\sqrt{D_1 t}}\phi (\frac{x^2}{D_1t}),
\label{app:eq:1D_dimensionanalysis}
\end{equation}
where $\phi$ is an arbitrary function to be determined.
If we plug eq. (\ref{app:eq:1D_dimensionanalysis}) into eq. (\ref{app:eq:1Ddiffusion}), the partial differential equation is reduced to an ordinary differential equation.
After some calculation, we obtain
\begin{equation}
N_1(x,t) = \frac{Q_0}{\sqrt{\pi}x_{d1}}e^{-x^2/x_{d1}^2},
\label{app:eq:1D1zone}
\end{equation}
where $x_{d1}=2\sqrt{D_1t}$ is the diffusion length scale. 

In the two zone diffusion case, based on dimensional analysis the solution should follow
\begin{equation}
N_2(x,t, x_b)=\frac{Q_0}{\sqrt{\pi}x_{d1}}\begin{cases}
Ae^{-x^2/x_{d1}^2} + Ee^{-(x-\alpha x_{b})^2/x_{d1}^2}, \, x<x_b,\\
Ce^{-x^2/x_{d2}^2} + Fe^{-(x-\beta x_{b})^2/x_{d2}^2} ,\, x\geq x_b.
\end{cases}
\label{app:eq:1Dformal_solution}
\end{equation}
where $x_{d2}=2\sqrt{D_2t}$.
$A, E, C, F, \alpha $ and $\beta$ are constants which are independent of $x$ and $t$. Our main task is to derive all the unknown constants. 


At first, we investigate the asymptotic behavior of $N_2(x,t,x_b)$. When $x\sim  x_{d1} \ll x_b$, the particle distribution is unaffected by the zone of $D_2$. It is expected that 
\begin{equation}
N_2(x,t,x_b)\rightarrow N_1 = \frac{Q_0}{\sqrt{\pi}x_{d1}}e^{-x^2/x_{d1}^2}, 
\end{equation}
which implies $A=1$.

Secondly, $N_2(x, t, x_b)$ is continuous at $x=x_{b}$ for arbitrary $D_1$, $D_2$ and $x_b$, which indicates
\begin{equation}
N_2(x_{b}^+,t, x_b)\propto N_2(x_{b}^-,t, x_b) \propto e^{-x_b^2/x_{d1}^2},
\end{equation}  
As a result, we find that $\alpha=2$, $C=0$, $\beta=1-\sqrt{D_2/D_1}$ and $F=1+E$. 

Now the solution becomes
\begin{equation}
N_2(x ,t, x_b)=\frac{Q_0}{\sqrt{\pi}x_{d1}}\\
\begin{cases}
e^{-x^2/x_{d1}^2} + Ee^{-(x-2x_{b})^2/x_{d1}^2}, \, x<x_b,\\
(1+E)e^{-[(x-x_{b})/x_{d2}+x_{b}/x_{d1}]^2}, \, x\geq x_b.
\end{cases}
\label{app:eq:1Dsolution}
\end{equation}
According to the conservation of particle, i.e.,  $\int_{-\infty}^{\infty} N_2(x,t,x_b)dx=Q_0$, we derive 
\begin{equation}
E=\frac{\sqrt{D_1}-\sqrt{D_2}}{\sqrt{D_1}+\sqrt{D_2}}.
\end{equation}

The analytical solution discussed here is consistent with that obtained in \cite{MI87} with a different method. The solution naturally preserves the continuity of diffusion flux at $x=x_{b}$, i.e.,
\begin{equation}
D_1\frac{\partial N_2(x,t,x_b)}{\partial x}\mid_{x=x_{b}^-}=D_2\frac{\partial N_2(x,t,x_b)}{\partial x}\mid_{x=x_{b}^+}.
\end{equation}
When $D_1=D_2$, the two zone solution simply recovers the single zone solution provided in eq. (\ref{app:eq:1D1zone}).

Next, we discuss the diffusion of particle in spherical symmetry. We denote subscript s1 and s2 for single zone and two zone diffusion in spherical symmetry respectively. The particle transport equation is 
\begin{equation}
\frac{\partial M(r, t)}{\partial t}=\frac{1}{r^2}\frac{\partial }{\partial r}\left(r^2 D(r)\frac{\partial M(r,t)}{\partial r}\right) +Q_0(E)\delta(\vec{r},t),
\label{app:eq:spherical_diffusion}
\end{equation}
where $M(r,t)$ is the spatial density of particle at radius $r$ and time $t$, $D(r)$ is the diffusion coefficient and $Q_0(E)$ is the injection constant. The particles are injected at $\vec{r}=0$ and $t=0$ as indicated by the delta function $\delta(\vec{r},t)$. The diffusion coefficient satisfies
\begin{equation}
D(r)=\begin{cases}
D_1, \, 0<r<r_{b},\\
D_2, \, r\geq r_{b},
\end{cases}
\end{equation}
where $r_b$ is the contact discontinuity. $D_1$ and $D_2$ are the diffusion coefficients for the two zone respectively. Eq. (\ref{app:eq:spherical_diffusion}) can be further simplified into
\begin{equation}
\frac{\partial [rM(r, t)]}{\partial t}=\frac{\partial }{\partial r}\left\lbrace D(r)\frac{\partial [rM(r,t)]}{\partial r}\right\rbrace +rQ_0(E)\delta(\vec{r},t)
\end{equation}
with $r\geq 0$. The solution for 1D diffusion in half space can be derived by subtracting the contribution from a source at $r=a$ and the flux from a image source at $r=-a$, see e.g., \cite{Chandrasekhar43}. In single zone case, it is shown that 
\begin{equation}
M_{s1}(r,t) =  \lim_{a\rightarrow 0}\frac{w_{s1}[N_1(r-a,t) -N_1(r+a,t)]}{r},
\end{equation}
where $w_{s1}$ is a constant determined by the conservation of particle, i.e., $\int 4\pi r^2 M_{s1}(r,t) dr=Q_0$.
After some calculation, it is found that 
\begin{equation}
M_{s1}(r,t) = \frac{Q_0}{\pi^{3/2}r_{d1}^3}e^{-r^2/r_{d1}^2},
\label{app:eq:1D1zone_spherical_solution}
\end{equation}
where $r_{d1}=2\sqrt{D_1t}$ is the diffusion length scale.

For two zone diffusion case, we can apply the same technique to solution in eq. (\ref{app:eq:1Dsolution}) and then obtain 
\begin{equation}
M_{s2}(r,t, r_b) =  \lim_{a\rightarrow 0}\frac{w_{s2}[N_2(r-a,t,r_b-a) -N_2(r+a,t, r_b+a)]}{r},
\end{equation}
where $w_{s2}$ is a constant determined by the conservation of particle. We went through the calculation and then derive
\begin{equation*}
M_{s2}(r,t, r_b) = \frac{ b(b+1)Q_0}{\pi^{3/2}r_{d1}^3[\rm 2b^2erf(r_b)-b(b-1)erf(2r_b)+2erfc(r_b)]}
\end{equation*}

\begin{eqnarray}
\times \begin{cases}
\left[e^{-r^2/r_{d1}^2} + E\left(\frac{2r_b}{r}-1\right)e^{-(r-2r_{b})^2/r_{d1}^2} \right], \, r<r_b,\\
(1+E)\left[\frac{r_b}{r}+b(1-\frac{r_b}{r})\right]e^{-[(r-r_{b})/r_{d2}+r_{b}/r_{d1}]^2}, \, r\geq r_b.\\
\end{cases}&&
\end{eqnarray}
where $r_{d2}=2\sqrt{D_2t}$, $b= \sqrt{D_1}/\sqrt{D_2}$ and
\begin{equation}
E=\frac{\sqrt{D_1}-\sqrt{D_2}}{\sqrt{D_1}+\sqrt{D_2}}.
\end{equation}
erf(x) is the error function and $\rm erfc(x) = 1-erf(x)$. The solution naturally preserves the continuity of diffusion flux at $r=r_{b}$, i.e.,
\begin{equation}
D_1\frac{\partial [r M_{2s}(r,t,r_b)]}{\partial r}\mid_{r=r_{b}^-}=D_2\frac{\partial [rM_{2s}(r,t,r_b)]}{\partial r}\mid_{r=r_{b}^+}.
\end{equation}
When $D_1=D_2$, the two zone solution simply recovers the single zone solution provided in eq. (\ref{app:eq:1D1zone_spherical_solution}).

\clearpage

\bsp	
\label{lastpage}
\end{document}